\documentclass[aps,prc,amsfonts,preprintnumbers,superscriptaddress,showpacs,nofootinbib]{revtex4}
\pdfoutput=1
\usepackage{graphics}
\usepackage{amsmath}
\usepackage{amssymb}
\usepackage{psfrag}
\usepackage{epsfig}
\usepackage{epsf}
\usepackage{float}
\usepackage{slashed}
\allowdisplaybreaks

\usepackage{color}

\newcommand{\fet}[1]{\mbox{\boldmath $#1$}}

\newcommand{\beq}{\begin{equation}}
\newcommand{\eeq}{\end{equation}}
\newcommand{\beqa}{\begin{eqnarray}}
\newcommand{\eeqa}{\end{eqnarray}}
\newcommand{\nn}{\nonumber \\ }

\newcommand{\pvec}[1]{\vec{#1}\mkern2mu\vphantom{#1}}

\numberwithin{equation}{section}
\renewcommand\theequation{\arabic{equation}}

\begin{document}

\title{Subleading contributions to the nuclear scalar isoscalar currents}

\author{H.~Krebs}
\email[]{Email: hermann.krebs@rub.de}
\affiliation{Ruhr-Universit\"at Bochum, Fakult\"at f\"ur Physik und
        Astronomie, Institut f\"ur Theoretische Physik II,  D-44780
        Bochum, Germany}
\author{E.~Epelbaum}
\email[]{Email: evgeny.epelbaum@rub.de}
\affiliation{Ruhr-Universit\"at Bochum, Fakult\"at f\"ur Physik und
        Astronomie, Institut f\"ur Theoretische Physik II,  D-44780
        Bochum, Germany}
\author{U.-G.~Mei{\ss}ner}
\email[]{Email: meissner@hiskp.uni-bonn.de}
\affiliation{Helmholtz-Institut~f\"{u}r~Strahlen-~und~Kernphysik~and~Bethe~Center~for~Theoretical~Physics,
~Universit\"{a}t~Bonn,~D-53115~Bonn,~Germany}
\affiliation{Institute~for~Advanced~Simulation,~Institut~f\"{u}r~Kernphysik,
and J\"{u}lich~Center~for~Hadron~Physics, Forschungszentrum~J\"{u}lich,~D-52425~J\"{u}lich,~Germany}
\affiliation{Tbilisi State University, 0186 Tbilisi, Georgia}
\date{\today}

\begin{abstract}
We extend our recent analyses of the nuclear vector, axial-vector and
pseudoscalar currents and derive the leading one-loop corrections to
the two-nucleon scalar current operator in the framework of chiral
effective field theory using the method of unitary transformation. 
We also show that the scalar current operators at zero momentum
transfer are directly related to the quark mass dependence of the
nuclear forces.   
\end{abstract}


\maketitle

\vspace{-0.2cm}

%%%%%%%%%%%%%%%%%%%%%%%%%%%%%%%%%%%%%%%%%%%%%%%%%%%%%%%%%%%%%%%%%%%%%%%%%%%%%%%%%
\section{Introduction}
\def\theequation{\arabic{section}.\arabic{equation}}
\label{sec:intro}

The first principles description of nuclei, nuclear matter and reactions is
one of the great challenges in contemporary physics with applications
ranging from low-energy searches for physics beyond the Standard Model (SM)
to properties of neutron stars and neutron star mergers. The currently
most efficient and feasible approach along this line relies on the
application of suitably taylored effective field theories (EFTs). In
particular, an extension of chiral perturbation theory to
multi-nucleon systems \cite{Weinberg:1990rz,Weinberg:1991um}, commonly referred to as chiral EFT, has
been applied over the last two decades to derive nuclear forces at
high orders in the EFT expansion in harmony with the spontaneously
broken approximate chiral symmetry of QCD
\cite{Epelbaum:2008ga,Machleidt:2011zz}. See Refs.~\cite{Reinert:2017usi,Entem:2017gor} for the
most accurate and precise chiral two-nucleon interactions at fifth
order and Refs.~\cite{Epelbaum:2019kcf,RodriguezEntem:2020jgp,Petschauer:2020urh,Ekstrom:2020slg,Piarulli:2020mop}
for a collection of review 
articles describing the current state-of-the-art in chiral EFT for
nuclear forces and selected applications. In parallel with these
developments, current operators describing the interactions of nuclear
systems with external vector, axial-vector and pseudoscalar sources
needed to study electroweak reactions driven by a single photon- or
$W$/$Z$-boson exchange have been worked out completely through fourth
order in the heavy-baryon formulation of chiral EFT with pions and
nucleons as the only dynamical degrees of freedom, see
Refs.~\cite{Park:1993jf,Park:1995pn} for the  pioneering studies by Park 
et al., Refs.~\cite{Kolling:2009iq,Kolling:2011mt,Krebs:2016rqz,Krebs:2019aka} for our calculations using
the method of unitary transformation \cite{Okubo:1954zz, Epelbaum:1998ka,Epelbaum:2010nr} and
Refs.~\cite{Pastore:2009is,Pastore:2011ip,Baroni:2015uza,Baroni:2016xll} for an
independent derivation by the Jlab-Pisa group in the framework of time-ordered
perturbation theory. A direct comparison of the expressions for the
current operators derived by different group is hindered by
their scheme dependence. However, at least for the
two-pion exchange axial-vector currents, our results \cite{Krebs:2016rqz}
appear to be not unitarily equivalent to the ones of the Pisa-Jlab group \cite{Baroni:2015uza}, see
Ref.~\cite{Krebs:2020rms} for a detailed discussion of the box diagram
contribution. We further emphasize that off-shell consistency of the
electroweak operators derived by our group
\cite{Kolling:2009iq,Kolling:2011mt,Krebs:2016rqz,Krebs:2019aka} and
the corresponding (unregularized) two- \cite{Epelbaum:2004fk,Epelbaum:2014efa} and three-nucleon
forces \cite{Bernard:2007sp,Bernard:2011zr} has been verified explicitly by means of the corresponding
continuity equations in Refs.~\cite{Krebs:2016rqz,Krebs:2019aka}.   

In this work we extend our earlier studies
\cite{Kolling:2009iq,Kolling:2011mt,Krebs:2016rqz,Krebs:2019aka} and
investigate the two-nucleon scalar current operators. Specifically, we
consider the two-flavor QCD Lagrangian in the presence of external vector,
axial-vector, scalar and pseudoscalar sources $v_\mu (x)$, $a_\mu (x)$, $s(x)$
and $p(x)$, respectively:
\beq
\mathcal{L} = \mathcal{L}_{\rm QCD}^0 + \bar q \gamma^\mu (v_\mu +
\gamma_5 a_\mu ) q - \bar q (s - i \gamma_5 p ) q\,,
\eeq
where $q$ denotes the doublet of the up and down quarks fields, while $\mathcal{L}_{\rm QCD}^0$
is the chirally invariant Lagrangian with massless up- and
down-quarks. Throughout this work, we employ the SU(2) formulation of
chiral EFT as done in our calculations of nuclear forces
\cite{Epelbaum:2014efa,Epelbaum:2014sza,Reinert:2017usi,Bernard:2007sp,Bernard:2011zr,Epelbaum:2006eu,Epelbaum:2007us,Krebs:2012yv,Krebs:2013kha}
and current operators
\cite{Kolling:2009iq,Kolling:2011mt,Krebs:2016rqz,Krebs:2019aka}. The
external sources are represented by Hermitian 2$\times$2 matrices in
the flavor space, and the original QCD Lagrangian is restored by
setting $v_\mu = a_\mu = p = 0$, $s = {\rm diag} (m_u, \, m_d)$. Here
and in what follows, we assume exact isospin symmetry with $m_u = m_d
\equiv m_q$. Embedded in the SM, the interactions between quarks and
the external vector and axial-vector sources are probed in electroweak
reactions involving hadrons or nuclei. Low-energy nuclear systems are nowadays 
commonly described by solving the many-body Schr\"odinger equation
with the nuclear forces derived in chiral EFT \cite{Epelbaum:2008ga,Machleidt:2011zz,Epelbaum:2019kcf}.
An extension to electroweak processes with nuclei requires the knowledge of the
corresponding nuclear current operators defined in terms of the
functional derivatives of the effective nuclear Hamiltonian in the
presence of external fields with respect to $v_\mu (x)$ and $a_\mu
(x)$ \cite{Krebs:2016rqz}. For the vector, axial-vector and pseudoscalar sources,
the corresponding expressions are already available up to fourth
chiral order
\cite{Kolling:2009iq,Kolling:2011mt,Krebs:2016rqz,Krebs:2019aka}. In
this work we focus on the response of nuclear 
systems to the external scalar source $s(x)$ and thus set  $v_\mu =
a_\mu = p = 0$. While the scalar currents cannot be probed
experimentally within the SM due to the absence of scalar
sources, they figure prominently in dark matter (DM) searches in a
wide variety of DM models such as e.g.~Higgs-portal DM and
weakly-interacting massive particles (WIMPs), see
\cite{Roszkowski:2017nbc,Kahlhoefer:2017dnp,Arcadi:2017kky} for recent review
articles. For example, the dominant interactions of a spin-1/2
Dirac-fermion DM particle $\chi$ with the strong sector of the SM is
given by the Lagrangian
\beq
\label{wimp}
\mathcal{L}_\chi = \bar \chi \chi \Big( \sum_i c_i m_i \,\bar q_i q_i +
  c_G \,\alpha_s G_{\mu \nu}^a G^{\mu \nu \, a} \Big)\,,
\eeq
where $i$ denotes the flavor quantum number, $G_{\mu \nu}^a$ is the
gluon field strength, $\alpha_s$ is the strong coupling constant
and the couplings $c_i$ ($c_G$) determine the strength of the
interaction between $\chi$ and quarks of flavor $i$ (gluons). Notice that the
contributions from coupling to heavy quarks (charm, bottom and top)
can be integrated out \cite{Shifman:1978zn} and the sum in
Eq.~(\ref{wimp}) can thus be taken only over the light quark flavors
by replacing the coupling constants $c_i$, $c_G$ with 
the corresponding effective ones. Thus, the scalar nuclear currents derived 
in our paper can be used to describe the interactions of nuclei with
DM particles emerging from their isoscalar coupling to the up- and
down-quarks $\propto (c_u + c_d)$.

Apart from their relevance for DM searches, the scalar currents are
intimately related to quark mass dependence of hadronic and nuclear
observables. For example, the pion-nucleon $\sigma$-term, $\sigma_{\pi N}$,
corresponds to the isoscalar scalar form factor of the
nucleon at zero momentum transfer times the quark mass and determines the amount of the
nucleon mass generated by the up- and down-quarks. Its value has
been accurately determined from the recent Roy-Steiner-equation analysis of
pion-nucleon scattering accompanied with pionic hydrogen and deuterium
data to be $\sigma_{\pi N} = (59.1 \pm 3.5)$~MeV \cite{Hoferichter:2015dsa}.  
For the status of lattice QCD calculations of $\sigma_{\pi N}$ see
Ref.~\cite{Aoki:2019cca}. As pointed out, however, in Ref.~\cite{Hoferichter:2016ocj},
there is relation between the $\sigma$-term and the S-wave $\pi N$ scattering lengths that
so far has not been checked for the lattice calculations.
Nuclear $\sigma$-terms and scalar form factors of light
nuclei have also been studied in lattice QCD, albeit presently at
unphysically large quark masses \cite{Beane:2013kca,Chang:2017eiq}. Interestingly, the scalar
matrix elements were found in these studies to be strongly affected by
nuclear effects (in contrast to the axial-vector and tensor charges), which
indicates that scalar exchange currents may play an important role. 
Last but not least, as will be shown below, the scalar isoscalar currents
are directly related to the quark mass dependence of the nuclear forces, a
subject that gained a lot of attention in the EFT community in connection
with ongoing lattice QCD efforts in the multibaryon sector
\cite{Epelbaum:2002gb,Beane:2002vs,Chen:2010yt,Soto:2011tb,Epelbaum:2013ij,Barnea:2013uqa,Baru:2015ira,Behrendt:2016nql,Baru:2016evv},
a conjectured infrared renormalization group limit cycle in QCD 
\cite{Braaten:2003eu,Epelbaum:2006jc}, searches for possible temporal variation of the light quark
masses \cite{Bedaque:2010hr,Berengut:2013nh}  and anthropic considerations related to the famous Hoyle state
in $^{12}$C
\cite{Epelbaum:2012iu, Epelbaum:2013wla,Meissner:2014pma,Lahde:2019yvr}.   

Clearly, nuclear scalar currents have already been studied before in
the framework of chiral EFT, see
e.g.~\cite{Prezeau:2003sv,Hill:2011be,Cirigliano:2012pq,Hoferichter:2015ipa,Hoferichter:2016nvd,Bishara:2016hek,Korber:2017ery,Hoferichter:2018acd}. For
the two-nucleon currents, only the dominant contribution at the chiral
order $Q^{-2}$ stemming from the one-pion exchange has been considered so far. Here and in what
follows, $Q \in \{ M_\pi/\Lambda_b, \, p/\Lambda_b \}$ denotes the
chiral expansion parameter, $M_\pi$ is the pion mass, $p$ refers to
the magnitude of three-momenta of external nucleons, while $\Lambda_b$ denotes  the
breakdown scale of the chiral expansion. For a detailed discussion of
the employed power counting scheme for nuclear currents see
Ref.~\cite{Krebs:2016rqz}. The two-body scalar current is suppressed
by just one power of the expansion parameter $Q$ relative to the
dominant one-body contribution. Such an enhancement relative to the
generally expected suppression of $(A+1)$-nucleon operators
relative to the dominant $A$-nucleon terms by $Q^2$ can be traced back to the
vertex structure of the effective Lagrangian and is not uncommon. For
example, one- and two-nucleon operators contribute at the same order
to the axial charge and electromagnetic current operators, see Table~II of Ref.~\cite{Krebs:2016rqz}
and Table~1 of Ref.~\cite{Krebs:2019aka}, respectively. For the scalar operator,
the relative enhancement of the two-body terms is caused by the
absence of one-body contributions at the expected leading order
$Q^{-4}$, see e.g.~Table~III of Ref.~\cite{Krebs:2016rqz} for the hierarchy
of the pseudoscalar currents. The first corrections to the scalar current appear at order $Q^{-2}$
from the leading one-loop diagrams involving a single-nucleon line
\cite{Cirigliano:2012pq}. In this paper we derive the subleading contributions to the
two-nucleon scalar isoscalar current operators at order $Q^0$. While
the one-body current is not yet available at the same accuracy
level, using empirical information on the scalar form factor of
the nucleon from lattice QCD instead of relying on its strict chiral
expansion may, in the future, provide a more reliable and efficient approach. A
similar strategy is, in fact, commonly used in studies of electromagnetic
processes, see e.g.~\cite{Phillips:2016arnps,Marcucci:2015rca} and
Ref.~\cite{Filin:2019eoe} for a recent example.

Our paper is organized as follows. In section~\ref{sec:2N}, we briefly
describe the derivation of the current operator using the method of
unitary transformation and provide explicit expressions for the
leading (i.e.~order-$Q^{-2}$) and subleading (i.e.~order-$Q^{0}$)
two-body contributions. Next, in section~\ref{sec:summaryCurrents}, we establish a
connection between the scalar currents at zero momentum transfer and
the quark mass dependence of the nuclear force. The obtained results
are briefly summarized in section~\ref{sec:summary}, while some further technical details
and the somewhat lengthy expressions for the two-pion exchange
contributions are provided in appendices~\ref{sec:appen} and \ref{sec:TPE}.

%%%%%%%%%%%%%%%%%%%%%%%%%%%%%%%%%%%%%%%%%%%%%%%%%%%%%%%%%%%%%%%%%%%%%%%%%%%%%%%%%
\section{Two-nucleon scalar operators}
\def\theequation{\arabic{section}.\arabic{equation}}
\label{sec:2N}

The derivation of the nuclear currents from the effective chiral
Lagrangian using the method of unitary transformation is described in
detail in Ref.~\cite{Krebs:2016rqz}. The explicit form of the effective Lagrangian
in the heavy-baryon formulation
\beq
\label{lagr}
\mathcal{L}_{\rm eff} = \mathcal{L}_{\pi}^{(2)} + 
\mathcal{L}_{\pi}^{(4)} + \mathcal{L}_{\pi N}^{(1)} +
\mathcal{L}_{\pi N}^{(2)} + \mathcal{L}_{\pi N}^{(3)} +
\mathcal{L}_{NN}^{(0)} + \mathcal{L}_{NN}^{(2)} 
\eeq
can be found in Refs.~\cite{Gasser:1987rb} and \cite{Fettes:2000gb} for the pionic and
pion-nucleon terms, respectively. The relevant terms in
$\mathcal{L}_{NN}$ will be specified in section~\ref{sec:cont}. As
already pointed out above, for the purpose of this study we switch off
all external sources except the scalar one, $s(x)$. To derive the
scalar currents consistent with the nuclear potentials in
Refs.~\cite{Epelbaum:2004fk,Epelbaum:2014efa,Bernard:2007sp,Bernard:2011zr,Epelbaum:2006eu,Epelbaum:2007us}
and 
electroweak currents in Refs.~\cite{Kolling:2009iq,Kolling:2011mt,Krebs:2016rqz,Krebs:2019aka}, we first
switch from the effective pion-nucleon Lagrangian to the corresponding Hamiltonian $H[s]$
using the canonical formalism and then apply the unitary
transformations $U_{\rm Okubo}$, $U_\eta$ and $U[s]$. Here and in what
follows, we adopt the notation of Ref.~\cite{Krebs:2016rqz}. In particular, the
Okubo  transformations $U_{\rm Okubo}$ \cite{Okubo:1954zz} is a ``minimal''
unitary transformation needed to derive nuclear forces by decoupling the purely nucleonic
subspace $\eta$ from the rest of the pion-nucleon Fock
space in the absence of external sources. However, as found in
Refs.~\cite{Epelbaum:2006eu}, the resulting nuclear potentials
$\eta U_{\rm Okubo}^\dagger H U_{\rm Okubo}^{} \eta$, 
with $\eta$ denoting the
projection operator onto the $\eta$-space, are non-renormalizable
starting from next-to-next-to-next-to-leading order (N$^3$LO)
$Q^4$.\footnote{The chiral expansion of the nuclear forces starts with
the order Q$^0$ (LO).}  To obtain renormalized nuclear potentials, a more general class of unitary
operators was employed in Refs.~\cite{Epelbaum:2006eu,Epelbaum:2007us} by performing additional
transformations $U_\eta$ on the $\eta$-space. The explicit form of the
``strong'' unitary operators $U_{\rm Okubo}$ and $U_\eta$ up to
next-to-next-to-leading order (N$^2$LO) can be found in
Refs.~\cite{Epelbaum:2006eu,Epelbaum:2007us,Bernard:2007sp,Bernard:2011zr}.
Nuclear currents can, in principle, be obtained by switching on the
external classical sources in the effective Lagrangian, performing the same
unitary transformations $U_{\rm Okubo} U_\eta$ as in the strong
sector, and taking functional derivatives with respect to the
external sources. However, similarly to the above mentioned
renormalization problem with the nuclear potentials, the current
operators obtained in this way can, in general, not be
renormalized. A renormalizable formulation of the current operators
requires the introduction of an even more general class of unitary transformation by
performing subsequent $\eta$-space rotations with the unitary
operators, whose generators depend on the external
sources. In Refs.~\cite{Krebs:2016rqz} and \cite{Krebs:2019aka}, such additional unitary
operators $U[a_\mu, \, p]$ and $U[v_\mu]$, subject to the constraints
$U[a_\mu , \, p ]_{a_\mu =p = 0} = U[v_\mu]_{v_\mu = 0} = \eta$,
are explicitly given up to N$^2$LO.  Notice that such 
unitary transformations are necessarily time-dependent through the
dependence of their generators on the external sources. This, in  general,
induces the dependence of the corresponding current operators on
the energy transfer and results in additional terms in the continuity
equations \cite{Krebs:2016rqz}. We now follow the same strategy for
the scalar currents and introduce additional $\eta$-space unitary
transformations $U[s]$, $U[s]_{s = m_q} = \eta$, in order to obtain
renormalizable currents. The most general form of the operator $U[s]$
at the chiral order we are working with is given in appendix~\ref{sec:appen} and
is parametrized in terms of four real phases $\alpha_i^s$, $i = 0, \ldots , 3$.
%{\bf EE: How can the generator $S_0^s$ give a nonvanishing
%  contribution? The vertex $H_{2,2}^{(2)}$ is antisymmetric in the
%  pion isospin indices, while $S_{0,2}^{(2)}$ is symmetric... But if
%  it doesn't generate any nonvanishing contribution, how can the
%  phase $\alpha_0^s$ be fixed as given below?}
The nuclear scalar current is defined via
\beqa
\label{current:def}
S(k)&=&\int d^4 x \,\exp\left( -i
  k\cdot x\right)\frac{\delta}{\delta
  s(x)}\bigg|_{s=m_q}\left[U^\dagger[s]U^\dagger_\eta U_{\rm Okubo}^\dagger
H[s] U_{\rm Okubo} U_\eta U[s] +\left( i\frac{\partial}{\partial
  t}U^\dagger[s]\right) U[s]\right],
\eeqa
see \cite{Krebs:2016rqz} for notation. 
%The renormalizability
%requirement of the scalar current is found to give two constraints on
%the unitary phases, namely:
%\beqa
%\label{standard:unitary:choise}
%\alpha_0^{s}=-1,\nn
%\alpha_1^{s}=\alpha_2^{s}+\alpha_3^{s}.
%\eeqa
While all the phases 
%$\alpha_{2,3}^{s}$
remain unfixed, they do not show
up in the resulting expressions for the nuclear current given in the
following sections. To the order we are
working, we therefore do not see any unitary ambiguity. 
%From now on we
%employ the choice of the unitary phases as given by
%Eq.~(\ref{standard:unitary:choise}). 

%%%%%%%%%%%%%%%%%%%%%%%%%%%%%%%%%%%%%
\subsection{Contributions at orders $Q^{-2}$}

\begin{figure}[tb]
\vskip 1 true cm
\includegraphics[width=0.065\textwidth,keepaspectratio,angle=0,clip]{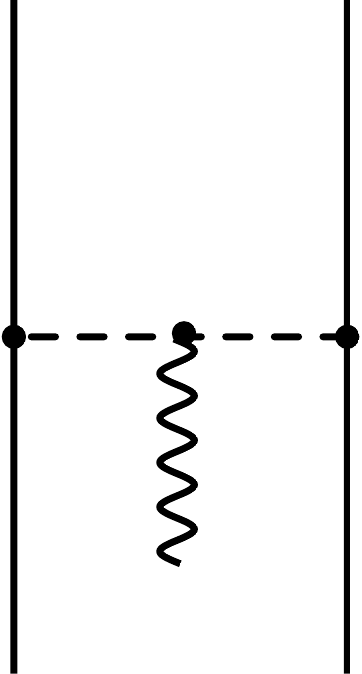}
    \caption{Diagram that leads to the dominant contribution of the
      2N scalar isoscalar current operator $S_{\rm
        2N}^{(Q^{-2})}$. Solid, dashed and wiggly lines denote
      nucleons, pions and external scalar sources, in order.
      Solid dots
      denote the leading-order vertices from the effective Lagrangians
      $\mathcal{L}_{\pi}^{(2)}$ and $\mathcal{L}_{\pi
        NN}^{(1)}$. 
\label{fig:tree} 
 }
\end{figure}

The chiral expansion of the 2N scalar isoscalar current starts at
order $Q^{-2}$. The dominant contribution is well known to emerge from
the one-pion exchange diagram shown in Fig.~\ref{fig:tree} and has the
form
\beqa
\label{SLO}
S_{\rm 2N}^{(Q^{-2})}&=&-\frac{g_A^2 M_\pi^2}{4 F_\pi^2 m_{q}}\frac{ \vec{q}_1\cdot \vec{\sigma}_1 
\vec{q}_{2}\cdot\vec{\sigma}_{2} }{ \left(M_\pi^2+q_1^2\right) 
\left(M_\pi^2+q_2^2\right)} {\fet \tau}_{1}\cdot{\fet 
\tau}_{2}\,,
\eeqa
where $g_A$ and $F_\pi$ are the nucleon axial-vector coupling and pion
decay constants, respectively, and $\vec q_i = \vec p_i^{\, \prime} - \vec p_i$ denotes the momentum
transfer of nucleon $i$. Further, $\vec \sigma_i$ ($\fet
\tau_i$) refer to the spin (isospin) Pauli matrices of nucleon
$i$. Here and in what follows, we follow the notation of our paper
\cite{Krebs:2016rqz}. In terms of the Fock-space operator $\hat S_{\rm 2N}$, the
expressions we give correspond to the matrix elements 
\beq
\langle \pvec p_1^{\, \prime}  \pvec p_2^{\, \prime} | \hat S_{\rm 2N} | \vec
p_1 \vec p_2 \rangle =: (2\pi)^{-3}\delta^{(3)} ( \pvec p_1^{\, \prime} + \pvec
p_2^{\, \prime} - \vec p_1
- \vec p_2 - \vec k\,  )   S_{\rm 2N} \,, 
\eeq
where $\vec p_i$  ($\vec p_i^{\, \prime}$) refers to the initial
(final) momentum of nucleon $i$, $\vec k$ is the momentum of the
external scalar source and the nucleon states are normalized according
to the nonrelativistic relation $\langle \vec p_i^{\, \prime} |  \vec
p_i \rangle = \delta^{(3)} (\vec p_i^{\, \prime} - \vec p \,)$. Finally,
we emphasize that the dependence of the scalar currents on $m_q$, which is
renormalization-scale dependent, reflects the fact
that in our convention, the external scalar source $s(x)$ couples to
the QCD density $\bar q q$ rather than $m_q \bar q q$. Thus, only the
combination $m_q \hat S_{\rm 2N} (k)$ is renormalization-scale independent.  
This is completely analogous to the pseudoscalar currents derived in
Ref.~\cite{Krebs:2016rqz}, and we refer the reader to that work for more details.

%%%%%%%%%%%%%%%%%%%%%%%%%%%%%%%%%%%%%
\subsection{One-pion-exchange contributions at order $Q^0$}
\label{sec:OnePionQ}

\begin{figure}[tb]
\vskip 1 true cm
%  \begin{center} 
%    \epsfxsize=4.3cm
%    \epsffile{fig2.eps}
\includegraphics[width=\textwidth,keepaspectratio,angle=0,clip]{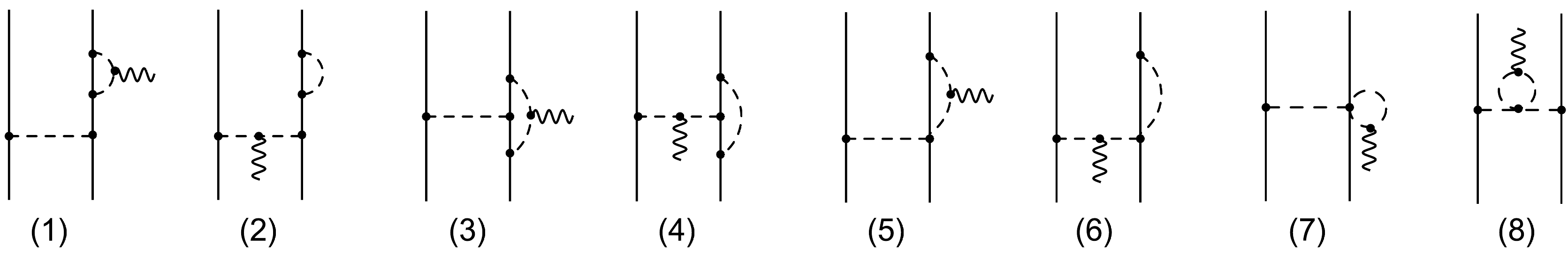}
    \caption{
         Non-tadpole one-loop one-pion-exchange diagrams contributing
         to $S_{\rm 2N}^{(Q^0)}$.  For notation, see Fig.~\ref{fig:tree}.
\label{fig:ope} 
 }
%  \end{center}
\end{figure}

\begin{figure}[tb]
\vskip 1 true cm
%  \begin{center} 
%    \epsfxsize=4.3cm
%    \epsffile{fig2.eps}
\includegraphics[width=15.4cm,keepaspectratio,angle=0,clip]{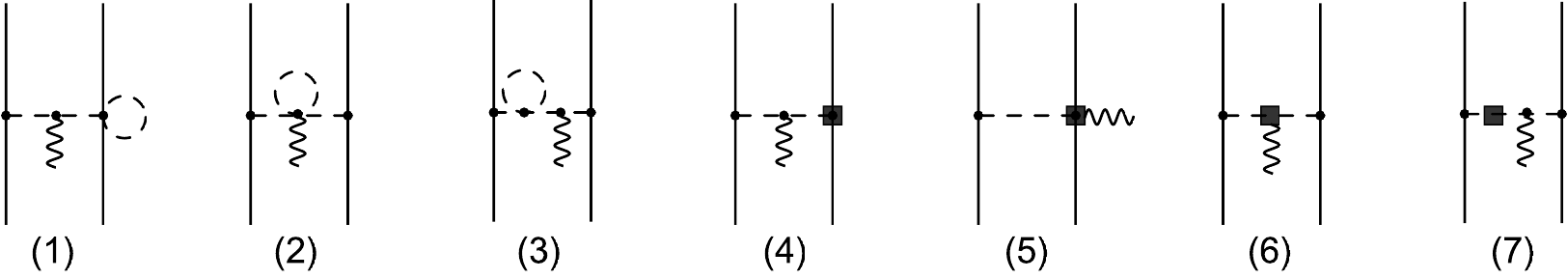}
    \caption{One-pion-exchange
      tadpole and tree-level diagrams that contribute to $S_{\rm 2N}^{(Q^0)}$. Filled squares denote
      the vertices from ${\cal  L}_{\pi N}^{(3)}$ and ${\cal  L}_{\pi}^{(4)}$ proportional to the
      low-energy constants $d_i$ and $l_i$, respectively. For remaining notation, see Fig.~\ref{fig:tree}.
\label{fig:tadpoles} 
 }
%  \end{center}
\end{figure}

Given that the first corrections to the pionic vertices are suppressed
by two powers of the expansion parameter and the absence of 
vertices in $\mathcal{L}_{\pi N}^{(2)}$ involving the scalar source and a
single pion, the first corrections to the two-nucleon current appear
at order $Q^0$. In Fig.~\ref{fig:ope} we show all one-loop one-pion-exchange diagrams
of non-tadpole type that contribute to the scalar current at this
order. Similarly, the corresponding tadpole and tree-level diagrams
yielding nonvanishing contributions are visualized in Fig.~\ref{fig:tadpoles}.  
It should be understood that the diagrams we show here and in what
follows do, in general, not correspond to Feynman graphs and serve for
the purpose of visualizing the corresponding types of contributions to the
operators. The meaning of the diagrams is specific to the method of
unitary transformation, see \cite{Krebs:2016rqz} for details. 
%We further emphasize that the contributions of diagrams (1) and (2) in
%Fig.~\ref{fig:ope} are affected by the additional unitary
%transformations $U [s]$. Only for the choice of phases given in
%Eq.~(\ref{standard:unitary:choise}) do the corresponding divergent
%contributions cancel against the counter terms in ${\cal  L}_{\pi N}$.
%Evaluating the contributions from the diagrams depicted in
%Figs.~\ref{fig:ope} and \ref{fig:tadpoles} for the choice of
%the unitary phases given in Eq.~(\ref{standard:unitary:choise}) 
Using dimensional regularization,
replacing all bare low-energy constants (LECs) $l_i$ and $d_i$ in terms of
their renormalized values $\bar l_i$ and $\bar d_i$ as defined in Eq.~(2.118) of \cite{Krebs:2016rqz}, and expressing
the results in terms of physical parameters $F_\pi$, $M_\pi$ and
$g_A$, see e.g.~\cite{Kolling:2011mt}, 
leads to our final result for the static order-$Q^0$ contributions to
the 2N one-pion-exchange scalar current operators:   
\beqa
\label{Current1pi}
S_{{\rm 2N:} \, 1\pi}^{(Q^0)}&=&
\frac{\vec{q}_1\cdot\vec{\sigma}_1}{q_1^2+M_\pi^2}\bigg[\vec{q}_2\cdot\vec{\sigma}_2
\bigg(\frac{o_1(k)}{q_2^2+M_\pi^2} +
o_2(k)\bigg)+\vec{k}\cdot\vec{\sigma}_2\Big(o_3(k)+q_2^2 o_4(k)\Big)\bigg]\; + \; 1
\leftrightarrow 2\,, 
\eeqa
where the scalar functions $o_i (k)$ are given by
%{\bf EE: I have slightly rewritten the expressions below. Please check. }
\bigskip
\beqa
o_1(k)&=&\frac{g_A M_\pi^2}{128\pi^2 F_\pi^4 m_q}
 \big[64\pi^2\bar{d}_{18}
      F_\pi^2 M_\pi^2+g_A k^2\bar{l}_4-g_A
    L(k)\left(2k^2+M_\pi^2
\right)+g_A\left(k^2+M_\pi^2\right)\big],\nn
o_2(k)&=&\frac{g_A M_\pi^2}{64\pi^2 F_\pi^4 m_q}
 \bigg[32\pi^2 F_\pi^2\left(2\bar{d}_{16} -\bar{d}_{18} \right)-g_A\bar{l}_4
%+g_A^3
-\frac{4g_A^3 L(k)\left(k^2+3 M_\pi^2\right)}{k^2+4M_\pi^2}
\bigg], \nn
o_3(k)&=&- \frac{g_A M_\pi^2 }{128\pi^2 F_\pi^4 k^2 m_q} \nn
&\times& \bigg[128\pi^2\bar{d}_{16} F_\pi^2 k^2+g_A^3\left(-k^2+M_\pi^2
\right)+2g_Ak^2
-\frac{4g_AL(k)}{k^2+4 M_\pi^2}\left(\left(2g_A^2+1
\right)k^4+\left(5g_A^2+4\right)k^2 M_\pi^2+g_A^2 M_\pi^4\right)
\bigg],\nn
o_4(k)&=&-\frac{g_A^4 M_\pi^2}{128\pi^2 F_\pi^4 k^2 m_q}
\frac{k^2+4 M_\pi^2(1-L(k))}{k^2+4 M_\pi^2}
\,, \label{OPEhiDefenition}
\eeqa
and the loop function $L(k)$ is defined as
\beq
L(k) = \frac{\sqrt{k^2 + 4 M_\pi^2}}{k} \ln \bigg(\frac{\sqrt{k^2 + 4
    M_\pi^2}+k}{2 M_\pi} \bigg) 
%\quad \quad 
%\mbox{and} \quad \quad 
%A(q) = \frac{1}{2 q} \arctan \bigg( \frac{q}{2 M_\pi} \bigg) 
\,.
\eeq

Finally, apart from the static contributions, we need to take
into account the leading relativistic corrections emerging from tree-level
diagrams with a single insertion of the $1/m$-vertices from the
Lagrangian $\mathcal{L}_{\pi N}^{(2)}$. Given our standard counting
scheme for the nucleon mass $m \sim \Lambda_b^2/M_\pi$, see
e.g.~\cite{Krebs:2016rqz}, these contributions are shifted from
the order $Q^{-1}$ to $Q^0$. However, the explicit evaluation of diagrams
emerging from a single insertion of the  $1/m$-vertices into the
one-pion-exchange graph in Fig.~\ref{fig:tree} leads to a vanishing
result. Given the relation between the scalar current operator and
the nuclear forces discussed in section~\ref{sec:summaryCurrents}, this observation is
consistent with the absence of relativistic corrections in the
(energy-independent formulation of the) nuclear forces at next-to-leading order. 

Last but not least, there are no contributions proportional to the
energy transfer $k_0$ which may appear from the explicit time dependence of the unitary
transformations in diagrams shown in Fig.~\ref{fig:ope}.

%%%%%%%%%%%%%%%%%%%%%%%%%%%%%%%%%%%%%
\subsection{Two-pion-exchange contributions}
\label{sec:2N2pi}

\begin{figure}[tb]
\vskip 1 true cm
\includegraphics[width=8.4cm,keepaspectratio,angle=0,clip]{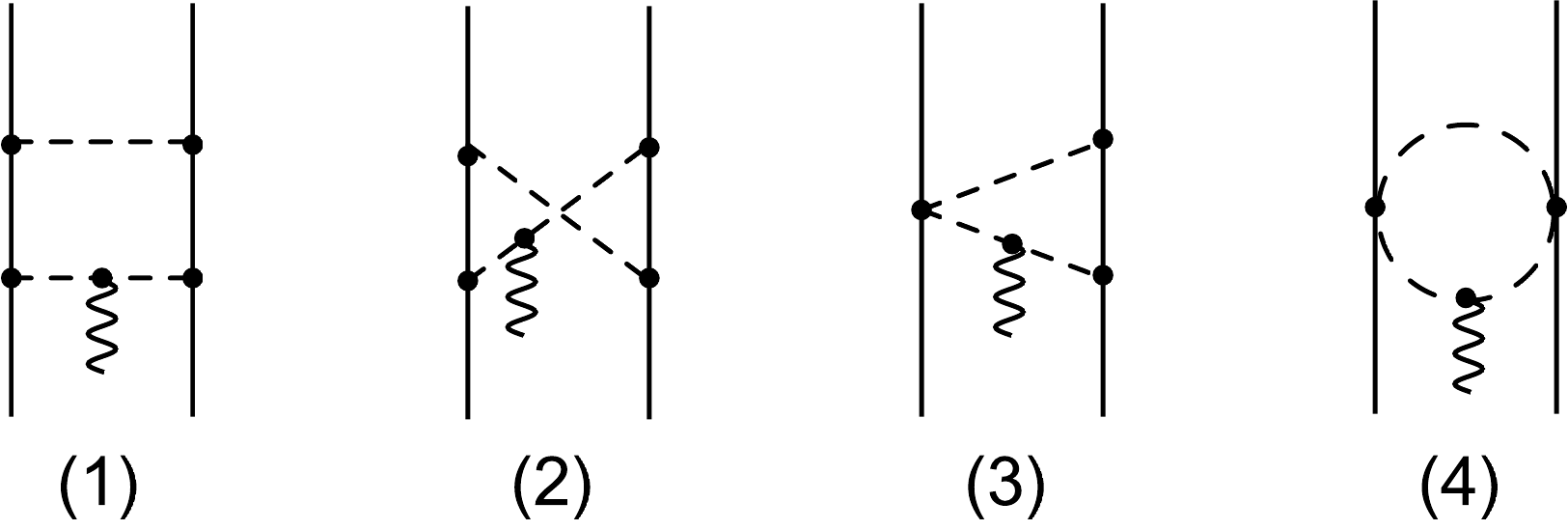}
    \caption{
         Two-pion-exchange diagrams contributing to $S_{\rm 2N}^{(Q^0)}$. For notation, see Fig.~\ref{fig:tree}.  
\label{fig:tpe} 
 }
%  \end{center}
\end{figure} 

We now turn to the two-pion exchange contributions. 
In Fig.~\ref{fig:tpe}, we show all diagrams yielding non-vanishing
results for the scalar current operator with two
exchanged pions. The final results for the two-pion exchange
operators read
\beqa
\label{Current2pi}
S_{{\rm 2N:} \, 2\pi}^{(Q^0)}= {\fet \tau}_1\cdot{\fet \tau}_2
\big[\vec{q}_1\cdot\vec{\sigma}_1\vec{k}\cdot\vec{\sigma}_2
t_1 + t_2\big] +
\vec{q}_1\cdot\vec{\sigma}_1\vec{q}_2\cdot\vec{\sigma}_2 t_3
+ \vec{q}_2\cdot\vec{\sigma}_1\vec{q}_1\cdot\vec{\sigma}_2
t_4 +\vec{q}_1\cdot\vec{\sigma}_1\vec{q}_1\cdot\vec{\sigma}_2 t_5
+\vec{\sigma}_1\cdot\vec{\sigma}_2 t_6   +  1
\leftrightarrow 2\,, 
\eeqa
where the scalar functions $t_i(k,q_1,q_2)$ are expressed in terms of
the three-point function. Their explicit form is given in appendix~\ref{sec:TPE}. Notice that
the (logarithmic) ultraviolet divergences in the two-pion
exchange contributions are absorbed into renormalization of the LECs
from $\mathcal{L}^{(2)}_{NN}$ described in the next section.

%%%%%%%%%%%%%%%%%%%%%%%%%%%%%%%%%%%%%
\subsection{Short-range contributions}
\label{sec:cont}

\begin{figure}[tb]
\vskip 1 true cm
\includegraphics[width=10.8cm,keepaspectratio,angle=0,clip]{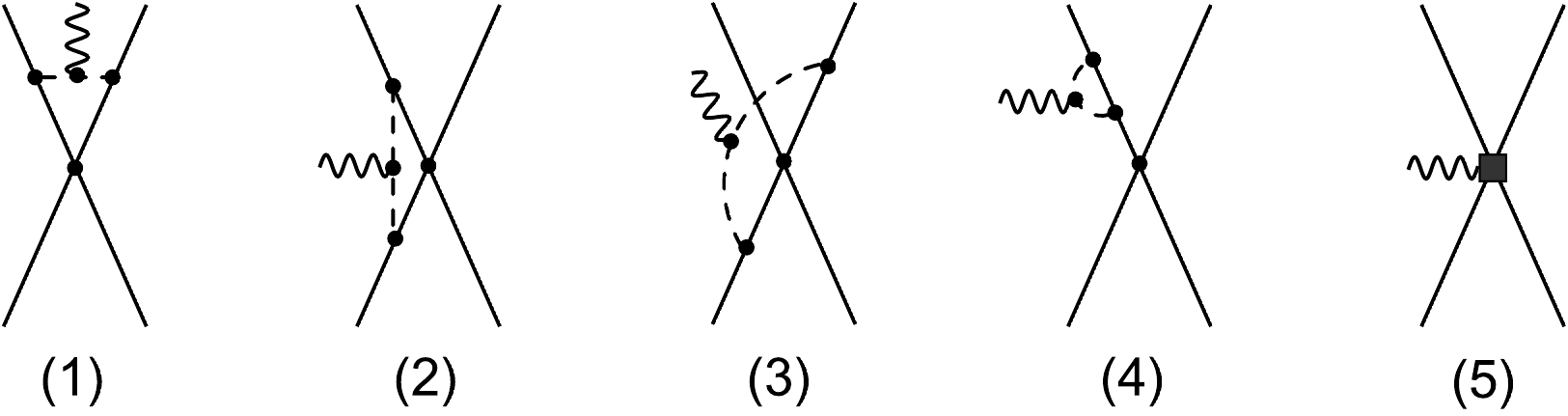}
    \caption{
         Loop diagrams with contact interactions contributing to $S_{\rm
           2N}^{ (Q^0)}$. Solid dots denote vertices from
        ${\cal L}_{\pi N}^{(1)}$, ${\cal L}_{\pi}^{(2)}$ or  ${\cal
          L}_{NN}^{(0)}$. Vertices from ${\cal
          L}_{NN}^{(2)}$ are denoted by filled squares. For remaining notation see Fig.~\ref{fig:tree}. 
\label{fig:contact} 
 }
%  \end{center}
\end{figure}

Finally, we turn to the contributions involving short-range
interactions. In Fig.~\ref{fig:contact}, we show all one-loop and
tree-level diagrams involving a single insertion of the contact
interactions that yield non-vanishing contributions to the scalar
current. The relevant terms in the effective Lagrangian have the form~\cite{Epelbaum:2007us,Epelbaum:2002gb} 
\beqa
\mathcal{L}_{NN}^{(0)}&=& - \frac{\overline{C}_S}{2} (N^\dagger N)^2 + 2 \overline{C}_T N^\dagger
S_\mu N N^\dagger S^\mu N,\nn
\mathcal{L}_{NN}^{(2)}&=& -\frac{D_S}{8} \langle \chi_+\rangle (N^\dagger N)^2 +
\frac{D_T}{2} \langle \chi_+\rangle N^\dagger
S_\mu N N^\dagger S^\mu N + \dots
\eeqa  
where $N$ is the heavy-baryon notation for the nucleon field with
velocity $v_\mu$, $S_\mu = - \gamma_5 [\gamma_\mu, \, \gamma_\nu ]
v^\nu /4 $ is the
covariant spin-operator, $\chi_+ = 2 B \left(u^\dagger (s + i p) u^\dagger + u
(s - i p) u\right)$, $B$, $\overline{C}_{S,T}$ and $D_{S,T}$ are
LECs\footnote{
 % As usual, we
%  write the Lagrangian in terms of the bare LECs, the renormalized versions are
%  given below.
  Since the symbols $C_{S,T}$ are commonly used to denote the $M_\pi$-dependent coefficients accompanying the
  momentum-independent contact operators in the NN potential, we
  follow here the convention of Ref.~\cite{Epelbaum:2002gb}  and use
  $\overline{C}_{S,T}$ to denote the
  corresponding bare LECs entering the effective Lagrangian.}, $\langle \ldots \rangle$ denotes the
trace in the flavor space, $u = \sqrt{U}$, and the 2$\times$2 matrix
$U$ collects the pion fields. Further, the ellipses refer to other terms that are not relevant for our
discussion of the scalar current operator. 

The total contribution of the diagrams of Fig.~\ref{fig:contact} can, after
renormalization, be written in the form
\beqa
\label{CurrentCont}
S_{\rm 2N: \, cont}^{ (Q^0)}&=& \vec{\sigma}_1\cdot\vec{\sigma}_2 s_1(k)
+ \vec{k}\cdot\vec{\sigma}_1\vec{k}\cdot\vec{\sigma}_2 s_2(k) + s_3(k)
\; + \; 1 \leftrightarrow 2\,,\label{ChargeContQTo1}
\eeqa
with the scalar functions $s_i(k)$ defined by 
%{\bf EE: I have slightly rewritten the expressions below. Please check. }
\beqa
s_1(k)&=&-\frac{M_\pi^2}{8\pi^2 F_\pi^2 m_q}
  \left[2 g_A^2 \overline{C}_T-4\pi^2 \bar D_T F_\pi^2
+\frac{g_A^2 \overline{C}_T L(k)\left(3k^2+4M_\pi^2\right)}{k^2+4M_\pi^2}
\right],\nn
s_2(k)&=&\frac{3 g_A^2
  \overline{C}_T M_\pi^2}{8\pi^2 F_\pi^2 k^2
m_q} \frac{k^2-4 M_\pi^2(L(k)-1)}{k^2+4M_\pi^2}, \nn
s_3(k)&=&\frac{M_\pi^2}{16\pi^2 F_\pi^2 m_q}
\left[g_A^2 \overline{C}_T+8\pi^2\bar D_S F_\pi^2
-\frac{2g_A^2 \overline{C}_T L(k)\left(3k^2+8M_\pi^2\right)}{k^2+4M_\pi^2}
\right]~.
\eeqa
The renormalized, scale-independent LECs $\bar D_{S}$,  $\bar D_T$ are
related to the bare ones $D_{S}$,  $D_T$  according to
\beqa
D_i=\bar D_i+\frac{\beta_i^{\rm NN}}{F^4}\lambda
+\frac{\beta_i^{\rm NN}}{16\pi^2 F^4}\ln\left(\frac{M_\pi}{\mu}\right),
\eeqa
with the corresponding $\beta$-functions given by
\beqa
\beta_S^{\rm NN}&=&\frac{1}{2}\left(1+6 g_A^2 - 15 g_A^4 + 24 F^2
  g_A^2 \overline{C}_T\right), \nn
\beta_T^{\rm NN}&=&\frac{1}{4}\left(1+6 g_A^2 - 15 g_A^4 + 48 F^2
  g_A^2 \overline{C}_T\right), 
\eeqa
and the quantity $\lambda$ defined as
\beqa
\lambda&=&\frac{\mu^{d-4}}{16\pi^2}\bigg(\frac{1}{d-4}+\frac{1}{2}\big(\gamma_E-\ln
4\pi -1\big)\bigg),
\eeqa
where $\gamma_E=-\Gamma^\prime(1)\simeq 0.577$ is the Euler constant, $d$ the number of space-time
dimensions and $\mu$ is the scale of dimensional regularization. 
Clearly, the $\overline{C}_T$-independent parts of the $\beta$-functions
emerge from the two-pion exchange contributions discussed in the
previous section. 

Notice that the LECs $\overline{C}_S$, $\overline{C}_T$, $\bar D_S$
and $\bar D_T$ also contribute to
the 2N potential. However, experimental data on nucleon-nucleon
scattering do not allow one to disentangle the $M_\pi$-dependence of
the contact interactions and only constrain the linear combinations of the LECs~\cite{Epelbaum:2002gb}
\beqa
C_S&=& \overline{C}_S+M_\pi^2\bar D_S, \quad C_T\,=\,
\overline{C}_T+M_\pi^2\bar D_T\;.
\eeqa
The LECs $\bar D_S$
and  $\bar D_T$ can, in principle, be determined once  reliable lattice QCD
results for two-nucleon observables such as e.g.~the $^3$S$_1$ and $^1$S$_0$ scattering lengths at
unphysical (but not too large) quark masses are available, see
Refs.~\cite{Lahde:2019yvr} and references therein for a discussion of
the current status of research along this line.

Last but not least, we found, similarly to the one-pion exchange
contributions, no $1/m$-corrections and no energy-dependent
short-range terms at the order we are working. Notice further that the
loop contributions to the contact interactions are numerically suppressed due to
the smallness of the LEC $C_T$ as a consequence of the approximate
SU(4) Wigner symmetry \cite{Wigner:1936dx,Mehen:1999qs}.

%%%%%%%%%%%%%%%%%%%%%%%%%%%%%%%%%%%%%%%%%%%%%%%%%%%%%%%%%%%%%%%%%%%%%%%%%%%%%%%%%
\section{Scalar current at zero momentum transfer}
\def\theequation{\arabic{section}.\arabic{equation}}
\label{sec:summaryCurrents}

If the four-momentum transfer $k_\mu$ of the scalar current is equal zero, one
can directly relate the current to the quark-mass derivative of the nuclear
Hamiltonian. To see this, we first rewrite the definition of the scalar current
in Eq.~(\ref{current:def}) in the form
\beqa
\label{current:at:k:equal:zero}
S(0)&=&\left[\left(\int d^4 x \frac{\delta}{\delta s(x)}\bigg|_{s=m_q}
  U^\dagger[s]\right), H_{\rm eff}\right] + U^\dagger_\eta U^\dagger_{\rm
  Okubo}\int d^4 x \frac{\delta H[s]}{\delta s(x)}\bigg|_{s=m_q} U_{\rm
  Okubo} U_\eta,
\eeqa
where the nuclear Hamiltonian $H_{\rm eff}$ is defined as 
\beqa
H_{\rm eff}&=&U_\eta^\dagger U_{\rm Okubo}^\dagger H[m_q] U_{\rm Okubo}^{} U_\eta^{}
\eeqa
and the unitary transformation $U[s]$ satisfies by construction 
\beqa
U[m_q]=1.
\eeqa 
Notice that the last term in the brackets in Eq.~(\ref{current:def})
vanishes for $k_0=0$.   On the other hand, we obtain
\beqa
\label{quark:derivative:force}
\frac{\partial H_{\rm eff}}{\partial m_q}
  &=&\left[\left(\frac{\partial}{\partial m_q}U_\eta^\dagger
    U_{\rm Okubo}^\dagger\right)U_{\rm Okubo}^{} U_\eta^{}, H_{\rm
    eff}\right] +  U^\dagger_\eta U^\dagger_{\rm Okubo}\frac{\partial H[m_q]}{\partial m_q}U_{\rm Okubo}^{} U_\eta^{}~.
\eeqa
Given the  trivial relation
\beqa
\int d^4 x \frac{\delta}{\delta s(x)}\bigg|_{s=m_q} H_{\rm eff} [s]
&=&\frac{\partial}{\partial m_q} H_{\rm eff} [m_q]\,,
\eeqa
the right-most terms in Eqs.~(\ref{current:at:k:equal:zero})
and (\ref{quark:derivative:force}) are equal, and we obtain the relation 
\beqa
\label{zero:momentum:current:rel}
S(0)&=&\frac{\partial H_{\rm eff}}{\partial
  m_q} + \left[\left(\int d^4 x \frac{\delta}{\delta s(x)}\bigg|_{s=m_q}
  U^\dagger[s]\right) - \left(\frac{\partial}{\partial m_q}U_\eta^\dagger
    U_{\rm Okubo}^\dagger\right)U_{\rm Okubo}^{} U_\eta^{}, H_{\rm eff}\right].
\eeqa
At the order we are working both commutators in this equation vanish
(independently on the choice of unitary phases)  leading to 
\beqa
\label{Smq}
S(0)&=&\frac{\partial H_{\rm eff}}{\partial
  m_q} + {\cal O}(Q^1).
\eeqa
In appendix~\ref{scalarcurrent:keq0} we demonstrate the validity of Eq.~(\ref{Smq}) for the
two-nucleon potential at NLO, see Ref.~\cite{Epelbaum:2002gb} for the calculation of the
quark mass dependence of nuclear forces using the method of unitary transformation.

It is important to emphasize that on the energy shell, i.e.~when
taking matrix elements in the eigenstates $| i \rangle$ and $| f \rangle$ of the
Hamiltonian  $H_{\rm eff}$ corresponding to the same energy, all
contributions from the commutator in Eq.~(\ref{zero:momentum:current:rel})
vanish leading to the exact relation 
\beqa
\label{ScalarCurrentAtZeroMomentum}
\langle f |S(0)| i \rangle &=& \bigg\langle f \bigg| \frac{\partial H_{\rm eff}}{\partial
  m_q} \bigg| i \bigg\rangle . 
  \eeqa
For eigenstates $| \Psi \rangle$ corresponding to a discrete energy
$E$, $H_{\rm
  eff} | \Psi \rangle = E  | \Psi \rangle $, the  Feynman-Hellmann
theorem allows one to interpret the scalar form factor at zero
momentum transfer in terms of the eigenenergy slope with respect to the quark mass:
\beqa
\label{ScalarCurrentDiscreteStates}
\langle \Psi|m_q S(0)| \Psi\rangle &=& m_q \frac{\partial E(m_q)}{\partial m_q}.
\eeqa
In particular, for $|\Psi\rangle$ being a single-nucleon state
at rest, the expectation value on left-hand side of
Eq.~(\ref{ScalarCurrentDiscreteStates}) is nothing but the pion-nucleon sigma-term
\beqa
\langle \Psi|m_q S(0)| \Psi\rangle \,=\, m_q \frac{\partial m_N(m_q)}{\partial
m_q} &\equiv & \sigma_{\pi N} \,,
\eeqa
and for an extension to resonances $|R\rangle$, see e.g.~Ref.~\cite{RuizdeElvira:2017aet}.

%%%%%%%%%%%%%%%%%%%%%%%%%%%%%%%%%%%%%%%%%%%%%%%%%%%%%%%%%%%%%%%%%%%%%%%%%%%%%%%%%
\section{Summary and conclusions}
\def\theequation{\arabic{section}.\arabic{equation}}
\label{sec:summary}

In this paper we have analyzed in detail the subleading contributions
to the nuclear scalar isoscalar current operators in
the framework of heavy-baryon chiral effective field theory. These
corrections are suppressed by two powers of the expansion parameter
$Q$ relative to the well-known leading-order contribution, see
Eq.~(\ref{SLO}). They comprise the one-loop corrections to the
one-pion-exchange and the lowest-order NN contact interactions as well
as the leading two-pion exchange contributions. No three- and more-nucleon
operators appear at the considered order. While the two-pion exchange
terms do not involve any unknown parameters, the one-pion exchange
contribution depends on a poorly known $\pi N$ LEC $\bar d_{16}$
related to the quark mass dependence of the nucleon axial coupling
$g_A$. It can, in principle, be determined from lattice QCD
simulations, see \cite{Chang:2018uxx,Alexandrou:2019brg} for some recent studies. The short-range part
of the scalar current depends on two unknown LECs which parametrize
the quark-mass dependence of the derivative-less NN  contact
interactions. In principle, these LECs can be extracted from the
quark-mass dependence of, say, the NN scattering length, see
Refs.~\cite{Epelbaum:2002gb,Beane:2002vs,Chen:2010yt,Epelbaum:2013ij,Baru:2015ira,Behrendt:2016nql,Baru:2016evv} for a related discussion.  Finally, we have explicitly demonstrated
that the scalar current operator at vanishing four-momentum transfer
is directly related to the quark-mass dependence of the nuclear force.  
The results obtained in our work are relevant for ongoing DM searches
and for matching to lattice QCD calculations in the few-nucleon
sector, see e.g.~\cite{Beane:2013kca,Chang:2017eiq} for recent studies
along this line.  

It is important to emphasize that our calculations are carried out
using dimensional regularization. For nuclear physics applications,
the obtained expressions for the scalar current operator need to be
regularized {\it consistently} with the nuclear forces, which
is a nontrivial task, see Refs.~\cite{Epelbaum:2019kcf,Krebs:2019uvm} for a discussion. Work along
these lines using the invariant higher derivative regularization \cite{Slavnov:1971aw} is in
progress.

%%%%%%%%%%%%%%%%%%%%%%%%%%%%%%%%%%%%%%%%%%%%%%%%%%%%%%%%%%%%%%%%%%%%%%%%%%%%%%%%%
\section*{Acknowledgments}

We are grateful to Martin Hoferichter and Jordy de Vries for sharing their insights into these topics.
This work was supported by DFG and NSFC through funds provided to the
Sino-German CRC 110 ``Symmetries and the Emergence of Structure in QCD" (NSFC
Grant No.~11621131001, DFG Grant No.~TRR110)
and BMBF (Grant No. 05P18PCFP1). 
The work of UGM was supported in part by VolkswagenStiftung (Grant no. 93562)
 and by the CAS President's International
 Fellowship Initiative (PIFI) (Grant No.~2018DM0034).

%%%%%%%%%%%%%%%%%%%%%%%%%%%%%%%%%%%%%%%%%%%%%%%%%%%%%%%%%%%%%%%%%%%%%%%%%%%%%%%%%
\appendix

%%%%%%%%%%%%%%%%%%%%%%%%%%%%%%%%%%%%%%%%%%%%%%%%%%%%%%%%%%%%%%%%%%%%%%%%%%%%%%%%%

%%%%%%%%%%%%%%%%%%%%%%%%%%%%%%%%%%%%%%%%%%%%%%%%%%%%%%%%%%%%%%%%%%%%%%%%%%%%%%%%%
\section{Additional unitary transformations}
\def\theequation{\Alph{section}.\arabic{equation}}
\setcounter{equation}{0}
\label{sec:appen}

At the order we are working, the general structure of the unitary
operator $U[s]$ can be written as
\beq
U[s] = \exp \bigg(\sum_{i=0}^{3}S_i^{s} - {\rm h.c.} \bigg)=1+\sum_{i=0}^{3}S_i^{s} -
{\rm h.c.}
+ \mathcal{O} \Big( \big( S_{i}^{s} \big)^2 \Big),
\eeq
where
\beqa
S_{0}^{s}&=&\alpha_{0}^{s} \eta  S_{0,2}^{(2)}  \lambda^2 \frac{1}{E_\pi^2}  H_{2,2}^{(2)} \eta,\nn
S_{1}^{s}&=&\alpha_{1}^{s} \eta  S_{0,2}^{(2)}  \lambda^2 \frac{1}{E_\pi^2}  H_{2,1}^{(1)} \lambda^1  \frac{1}{E_\pi}  
H_{2,1}^{(1)}  \eta,\nn
S_{2}^{s}&=&\alpha_{2}^{s} \eta  S_{0,2}^{(2)}  \lambda^2 \frac{1}{E_\pi}  H_{2,1}^{(1)} \lambda^1  \frac{1}{E_\pi^2}  
H_{2,1}^{(1)}  \eta,\nn
S_{3}^{s}&=&\alpha_{3}^{s} \eta  H_{2,1}^{(1)} \lambda^1  \frac{1}{E_\pi}   S_{0,2}^{(2)}  \lambda^1 \frac{1}{E_\pi^2}  
H_{2,1}^{(1)} \eta. \label{add:unitary:transf}
\eeqa
Here and in what follows, we use the notation of
Ref.~\cite{Krebs:2016rqz}. Furthermore, $S_{n,p}^{(\kappa)}$
denotes an interaction from the Hamiltonian with a single insertion of
the scalar current $s(x)-m_q$~\footnote{Note that the forces and
  currents are calculated at $s(x)=m_q$. In order to ensure the
  restriction $U[s\equiv m_q]=1$ for the employed additional unitary
  transformations, the interaction operator
  $S_{0,2}^{(2)} $ has to be proportional $s(x)-m_q$.}, $n$ nucleon and $p$ pion fields. The
superscripts $\kappa$ refer to the inverse mass dimension
of the corresponding coupling constant given by 
\beqa
\kappa&=&d + \frac{3}{2}n + p + c_v +c_a + 2 c_p +2 c_s - 4 \,,
\eeqa
where $d$, $n$ and $p$ denote the number of derivatives or pion mass
insertions  at a given vertex, number of nucleon  and pion fields, respectively.  Further,
$c_v$, $c_a$, $c_p$ and $c_s$ refer to the number  
of external vector, axial-vector, pseudoscalar and scalar sources, in
order.

%%%%%%%%%%%%%%%%%%%%%%%%%%%%%%%%%%%%%%%%%%%%%%%%%%%%%%%%%%%%%%%%%%%%%%%%%%%%%%%%%
\section{Two-pion exchange contributions to the scalar current}
\label{sec:TPE}

The scalar functions $t_i (q_1,q_2,k)$, $i=1, \ldots , 6$, with $q_i
\equiv | \vec q_i|$ and $k \equiv |\vec k|$ entering the expression
(\ref{Current2pi})  for the two-pion exchange current are given
by
\beqa
m_q
t_1&=&\frac{g_A^4M_\pi^2}{128\pi^2F_\pi^4k^2}-\frac{g_A^4M_\pi^4L(k)}{32
\pi^2F_\pi^4k^2\left(k^2+4M_\pi^2\right)},\nn
m_q t_2&=&
-\frac{\left(g_A^2-1\right) M_\pi^2}{8F_\pi^4}
\left(\frac{\left(g_A^2-1\right)k^2q_1^2
q_2^2}{4\left((\vec{q}_ 1\cdot\vec{q}_ 2)^2-q_1^2q_2^2
\right)}+\left(3g_A^2+1\right)M_\pi^2+2g_A^2q_1^2\right)i\,I(4;0,1;q_1,1;k,1;0,0)\nn
&-&\frac{M_\pi^2
L(q_1)}{256\pi^2F_\pi^4\left((\vec{q}_1\cdot\vec{q}_2)^2-q_1^2
q_2^2\right)}\left(-\frac{g_A^4}{\left(4M_\pi^2+q_1^2\right)\left(q_1^2
q_2^2\left(k^2+4M_\pi^2\right)-4M_\pi^2(\vec{q}_1\cdot\vec{q}_ 2)^2
\right)}\Big(2k^8\left(8M_\pi^4+3M_\pi^2q_1^2
\right)\right. \nn
&-&\left. k^6\left(M_\pi^4\left(52q_1^2+64q_2^2
\right)+M_\pi^2\left(19q_1^4+36q_1^2q_2^2\right)+4q_1^4q_2^2
\right)+k^4\left(M_\pi^4\left(60q_1^4+52q_1^2q_2^2+96q_2^4
\right)\right.\right.\nn
&+&\left.\left. M_\pi^2\left(21q_1^6+35q_1^4q_2^2+64q_1^2q_2^4\right)+
5q_1^4q_2^2\left(q_1^2+2q_2^2\right)\right)+k^2
Q_{-}^2\left(M_\pi^4\left(-28q_1^4+12q_1^2q_2^2+64q_2^4
\right)\right.\right.  \nn
&+&\left.\left. M_\pi^2\left(-9q_1^6+5q_1^4q_2^2+44q_1^2q_2^4\right)+
q_1^4q_2^2\left(q_1^2+8q_2^2\right)\right)+
Q_{-}^6\left(4M_\pi^4\left(q_1^2-4q_2^2\right)+M_\pi^2\left(
q_1^4-10q_1^2q_2^2\right)-2q_1^4q_2^2\right)
\Big)\right. \nn
&-&\left. 2g_A^2\left(k^4-k^2\left(q_1^2+2q_2^2\right)-q_2^2Q_{-}^2
\right)+q_1^2\left(k^2-Q_{-}^2\right)
\right)\nn
&-&\frac{M_\pi^2
L(k)}{512\pi^2 F_\pi^4 \left((\vec{q}_1\cdot\vec{q}_2)^2-q_1^2
q_2^2\right)}\bigg(-\frac{g_A^4}{\left(k^2+4M_\pi^2\right)\left(q_1^2q_2^2\left(k^2+4M_\pi^2
\right)-4M_\pi^2(\vec{q}_1\cdot\vec{q}_2)^2
\right)}\Big(5k^{10}M_\pi^2\nn
&+& k^8\left(20M_\pi^4-46
M_\pi^2q_1^2-7q_1^2q_2^2\right)+2k^6
q_1^2\left(-92M_\pi^4+M_\pi^2\left(37q_1^2+q_2^2\right)+15q_1^2
q_2^2\right)+2k^4\left(52M_\pi^4q_2^2\left(q_1^2+3q_2^2
\right)\right.  \nn
&+&\left.  M_\pi^2\left(83q_1^4q_2^2-23q_1^6\right)-8q_1^6q_2^2+8
q_1^4q_2^4\right)-4k^2M_\pi^2q_1^4Q_{-}^2\left(58M_\pi^2-
q_1^2+27q_2^2\right)+4M_\pi^2q_1^4Q_{-}^2\left(16M_\pi^2\left(
q_1^2-3q_2^2\right)\right. \nn
&+&\left. q_1^4-2q_1^2q_2^2\right)
\Big)+8g_A^2\left(2q_1^2q_2^2+q_2^2
\vec{q}_1\cdot\vec{q}_2-(\vec{q}_1\cdot\vec{q}_2)^2\right)-2k^2\vec{q}_1\cdot\vec{q}_2
\bigg)-
\frac{\left(g_A^2+1\right)^2M_\pi^2}{128\pi^2F_\pi^4}\;,\nn
m_q t_3&=&\frac{3 i\, g_A^4\,I(4;0,1;q_1,1;k,1;0,0)M_\pi^2(\vec{q}_1\cdot\vec{q}_2)
^2}{8F_\pi^4\left(q_1^2q_2^2-(\vec{q}_1\cdot\vec{q}_2)^2\right)}+\frac{3g_A^4M_\pi^2q_1^2q_2^2\left(k^2+Q_{-}^2
\right)L(q_1)\vec{q}_1\cdot\vec{q}_2}{64\pi^2F_\pi^4\left(q_1^2q_2^2-(\vec{q}_1\cdot\vec{q}_2)^2\right)\left(q_1^2q_2^2\left(k^2
+4M_\pi^2\right)-4M_\pi^2(\vec{q}_1\cdot\vec{q}_2)^2\right)}\nn
&+&\frac{3g_A^4M_\pi^2L(k)}{64\pi^2F_\pi^4}\left(\frac{1}{k^2+4M_\pi^2}-
\frac{q_1^2q_2^2\left(k^4-Q_{-}^4\right)}{4\left(q_1^2q_2^2-(\vec{q}_1\cdot\vec{q}_2)^2
\right)\left(q_1^2q_2^2\left(k^2+4M_\pi^2\right)-4M_\pi^2(\vec{q}_1\cdot\vec{q}_2)^2\right)}
\right)\;,\nn
m_q
t_4&=&m_q t_3+\frac{3i\,g_A^4\,I(4;0,1;q_1,1;k,1;0,0)M_\pi^2(q_1^2q_2^2-(\vec{q}_1\cdot\vec{q}_2)^2)}{8F_\pi^4
\left(q_1^2q_2^2-(\vec{q}_1\cdot\vec{q}_2)^2\right)}\;,\nn
m_q
t_5&=&
\frac{3g_A^4M_\pi^2q_2^2L(k)\left(-k^6 M_\pi^2+k^4
\left(3M_\pi^2 Q_+^2+2q_1^2q_2^2
\right)-3k^2M_\pi^2Q_{-}^4+M_\pi^2Q_{-}^4Q_{+}^2\right)}{64\pi^2F_\pi^4
\left(k^2+4M_\pi^2\right)\left(q_1^2q_2^2-(\vec{q}_1\cdot\vec{q}_2)^2
\right)\left(q_1^2q_2^2\left(k^2+4M_\pi^2\right)-4M_\pi^2(\vec{q}_1\cdot\vec{q}_2)^2\right)}\nn
&-&\frac{3g_A^4M_\pi^2q_1^2q_2^4\left(k^2-Q_{-}^2
\right)L(q_2)}{64\pi^2F_\pi^4\left(q_1^2q_2^2-(\vec{q}_1\cdot\vec{q}_2)^2
\right)\left(q_1^2q_2^2\left(k^2+4M_\pi^2\right)-4M_\pi^2(\vec{q}_1\cdot\vec{q}_2)^2
\right)}-\frac{3i\,g_A^4\,I(4;0,1;q_1,1;k,1;0,0)M_\pi^2q_2^2\vec{q}_1\cdot\vec{q}_2}{4F_\pi^4\left(q_1^2q_2^2-(\vec{q}_1\cdot\vec{q}_2)^2
\right)}\nn
&-&\frac{3g_A^4M_\pi^2q_1^2q_2^4\left(k^2+Q_{-}^2
\right)L(q_1)}{64\pi^2F_\pi^4\left(q_1^2q_2^2-(\vec{q}_1\cdot\vec{q}_2)^2\right)
\left(q_1^2q_2^2\left(k^2+4M_\pi^2\right)-4M_\pi^2(\vec{q}_1\cdot\vec{q}_2)^2
\right)}\;,\nn
m_q t_6&=&-\frac{3g_A^4M_\pi^2 L(k)\left(-k^6 M_\pi^2+k^4
\left(3M_\pi^2 Q_+^2+2q_1^2q_2^2
\right)-3k^2M_\pi^2Q_{-}^4+M_\pi^2Q_{-}^4Q_{+}^2\right)}{128\pi^2F_\pi^4\left(k^2+4M_\pi^2\right)\left(q_1^2q_2^2\left(k^2+4M_\pi^2\right)-4M_\pi^2(
\vec{q}_1\cdot\vec{q}_2)^2\right)}\nn
&+&\frac{3g_A^4M_\pi^2q_1^2q_2^2\left(k^2+Q_{-}^2
\right)L(q_1)}{64\pi^2F_\pi^4\left(q_1^2q_2^2\left(k^2+4M_\pi^2
\right)-4M_\pi^2(\vec{q}_1\cdot\vec{q}_2)^2
\right)}\;,
\label{giOfq1TPEStrF}
\eeqa
where $Q_\pm^2 \equiv q_1^2 \pm q_2^2$. 
Here, the scalar function 
$I(d:p_1,\nu_1;p_2,\nu_2;p_3,\nu_3;0,\nu_4)$ of the four-momenta $p_i$ is defined in terms of the integrals
\beqa
I(d:p_1,\nu_1;p_2,\nu_2;p_3,\nu_3; 0,\nu_4)&=&\int
\frac{d^d l}{(2\pi)^d}\prod_{j=1}^3\frac{1}{[(l+p_j)^2-M_\pi^2+i \epsilon]^{\nu_j}}
\frac{1}{[v\cdot l+i \epsilon]^{\nu_4}}\,.
\quad\quad
\eeqa
For the
case at hand with $p_i^0 =0$ and $\nu_1=\nu_2=\nu_3=1$ and $\nu_4=0$,
it is a standard three-point function with only pionic
propagators. Its explicit form is
given by
\beqa
I(4:0,1;q_1,1;k,1; 0,0)&=&\frac{i}{16\pi^2} \int_0^1d t\int_0^t dy \frac{1}{C}\frac{1}{(y-y_1)(y-y_2)},
\eeqa
with
\beqa
y_1&=&\frac{D}{2C}+\sqrt{\frac{D^2+4A C}{4 C^2}},\quad \quad
y_2\,=\,\frac{D}{2C}-\sqrt{\frac{D^2+4A C}{4 C^2}},
\eeqa
and $A\,=\,M^2+q_1^2(1-t)t$, $B\,=\,-2\vec{q}_1\cdot\vec{q}_2$,
$C\,=\,q_2^2$ and $D\,=\, 2\vec{q}_1\cdot\vec{q}_2+q_2^2 + t B$.
For $\vec{k}=0$, the three-point function reduces to a two-point
function
\beqa
I(4:0,1;q_1,1;k,1;
0,0)\big|_{\vec{k}=0}&=&-\frac{i}{8\pi^2}\frac{L(q_1)}{4 M_\pi^2+q_1^2} \;.\label{three_point_f_at_0}
\eeqa
Two-pion-exchange contribution to the scalar current reduces in this
case to
\beqa
m_q S_{\rm 2N: 2\pi}^{(Q^0)}\big|_{\vec{k}=0}&=&
\frac{M_\pi^2}{64\pi^2F_\pi^4\left(4M_\pi^2
+q_1^2\right)}\bigg[\frac{L(q_1)}{4M_\pi^2+q_1^2}\Big(6g_A^4\left(4M_\pi^2+q_1^2\right)\left(q_1^2
\vec{\sigma}_1\cdot\vec{\sigma}_2-q_1\cdot\vec{\sigma}_1q_1\cdot\vec{
\sigma}_2\right)\nn
&+&\left(16 M_\pi^4\left(-8g_A^4+4g_A^2+1
\right)+8 M_\pi^2q_1^2\left(-10g_A^4+5g_A^2+1\right)+q_1^4\left(-11g_A^4
+6g_A^2+1\right)\right)\fet\tau_1 \cdot\fet\tau_2 \Big)\nn
&-&\frac{1}{2}\left(4 M_\pi^2\left(15 g_A^4-2g_A^2+1
\right)+q_1^2\left(1-2 g_A^2+17 g_A^4\right)\right)\fet \tau_1 \cdot\fet
\tau_2\bigg]\;.
\eeqa
We will use this expression in appendix~\ref{scalarcurrent:keq0} to demonstrate the validity of
Eq.~(\ref{Smq}).

\section{Scalar current at zero momentum transfer}
\label{scalarcurrent:keq0}
In this appendix we demonstrate the validity of
Eq.~(\ref{Smq}). The quark mass dependence of the NLO nuclear force has
been extensively discussed in~\cite{Epelbaum:2002gb}. The explicit
expressions for effective potential at NLO
\beqa
V^{\rm OPE} + V^{\rm TPE} + V^{\rm cont}
\eeqa
can be found in Eqs.~(2.82), (2.83) and (2.84) of that paper. The authors
of~\cite{Epelbaum:2002gb} used the unitary transformation technique to
derive the nuclear force. Due to the appearance of the
time-derivative-dependent Weinberg-Tomozawa interaction, there appears an
additional derivativeless two-pion-four-nucleon-field vertex in the
Hamiltonian \cite{Epelbaum:2007us}, that  
leads to the tadpole diagram shown in Fig.~\ref{fig:tadpole_nn} which
was not considered in Ref.~\cite{Epelbaum:2002gb}. It generates 
% This
% diagram gives
an additional logarithmic contribution:

\begin{figure}[tb]
\vskip 1 true cm
\includegraphics[width=0.15\textwidth,keepaspectratio,angle=0,clip]{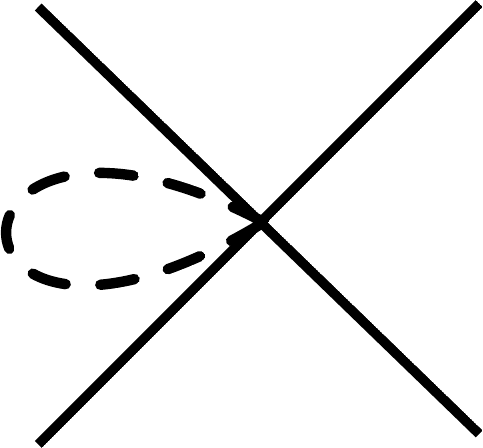}
    \caption{Short-range tadpole diagram which gives an additional
      contribution in the
      Hamiltonian formalism. 
\label{fig:tadpole_nn} 
 }
\end{figure}

\beqa
\delta V^{\rm cont}&=&
\frac{\fet{\tau}_1\cdot\fet{\tau}_2 \widetilde{M}_\pi^2}{64 F_\pi^4
  \pi^2} \ln\left(\frac{\widetilde{M}_\pi}{M_\pi}\right). 
\label{delta_V_tadpole_NN}
\eeqa
Here, we use the same notation as in Ref.~\cite{Epelbaum:2002gb} with
$\widetilde{M}_\pi$ denoting the pion mass at an unphysical quark mass value
and $M_\pi$ denoting  the physical pion mass.
At NLO, we have 
\beqa
V_{\rm NLO}&=&V^{\rm OPE} + V^{\rm TPE} + V^{\rm cont} + \delta V^{\rm cont}.
\eeqa
Taking derivative of the nuclear force in the quark mass is equivalent
to taking derivative in $\widetilde{M}_\pi^2$,
\beqa
\frac{\partial V_{\rm NLO}}{\partial m_q}&=&2 B \frac{\partial V_{\rm NLO}}{\partial \widetilde{M}_\pi^2}\bigg|_{\widetilde{M}_\pi=M_\pi}\,=\, \frac{M_\pi^2}{m_q}\frac{\partial V_{\rm NLO}}{\partial \widetilde{M}_\pi^2}\bigg|_{\widetilde{M}_\pi=M_\pi}.
\eeqa
Applying this operator to Eqs.~(2.82), (2.83) and (2.84)
of~\cite{Epelbaum:2002gb} and to Eq.~(\ref{delta_V_tadpole_NN}) of
that paper we obtain
\beqa
\frac{\partial V^{\rm OPE}}{\partial
  m_q}&=&\fet{\tau}_1\cdot\fet{\tau}_2\vec{\sigma}_1\cdot\vec{q}\,\vec{\sigma}_2\cdot\vec{q}\,\frac{M_\pi^2
  g_A}{4 m_q F_\pi^2}\bigg(\frac{g_A - 4 \bar{d}_{18}
  M_\pi^2}{(q^2+M_\pi^2)^2}\nn
&+&\frac{1}{8 F_\pi^2 \pi^2(q^2+M_\pi^2)}\Big(3
g_A^3+ g_A \bar{l}_4+32
F_\pi^2\pi^2\big(\bar{d}_{18}-2\bar{d}_{16} \big)\Big)\bigg)\;,\nn
\frac{\partial}{\partial
  m_q}\Big(V^{\rm TPE}+V^{\rm cont}+\delta V^{\rm
  cont}\Big)&=&\frac{M_\pi^2 L(q)}{4 m_q F_\pi^4\pi^2}\bigg(\frac{g_A^4
  M_\pi^4\fet{\tau}_1\cdot\fet{\tau}_2}{(q^2+4M_\pi^2)^2}+\frac{g_A^2}{8(q^2+4
  M_\pi^2)}\big(4(g_A^2-1) M_\pi^2\fet{\tau}_1\cdot\fet{\tau}_2\nn
&-&3
g_A^2(\vec{\sigma}_1\cdot\vec{q}\,\vec{\sigma}_2\cdot\vec{q}+4
M_\pi^2\vec{\sigma}_1\cdot\vec{\sigma}_2)\big)+\frac{1}{16}\big(6 g_A^4\vec{\sigma}_1\cdot\vec{\sigma}_2+(1+6g_A^2-11
g_A^4)\fet{\tau}_1\cdot\fet{\tau}_2\big)\bigg)\nn
&+&\frac{g_A^4
  M_\pi^4\fet{\tau}_1\cdot\fet{\tau}_2}{16 m_q F_\pi^4\pi^2 (q^2+4
  M_\pi^2)} + \frac{M_\pi^2}{384 m_q F_\pi^4\pi^2}\Big(384 F_\pi^4 \pi^2
\bar D_S+70 g_A^4 -4 g_A^2(36 F_\pi^2 \overline{C}_T + 5)\nn
&-&2 
+
\vec{\sigma}_1\cdot\vec{\sigma}_2(384 F_\pi^4\pi^2 \bar D_T+35
g_A^4 - 2 g_A^2(5+72 F_\pi^2 \overline{C}_T)-1)\Big)\;.
\eeqa
It is important to emphasize that in Ref.~\cite{Epelbaum:2002gb}, the short-range
LECs $\bar D_S$ and $\bar D_T$ have been shifted to absorb
all momentum-independent contributions generated by the
two-pion-exchange.  The corresponding shifts
for $\bar D_S$ and $\bar D_T$ are given by
\beqa
\bar D_S&\to&\bar D_S -\frac{1+g_A^2+4 g_A^4}{48
  F_\pi^4\pi^2}, \quad \quad \bar D_T\,\to\, \bar D_T -
\frac{1+g_A^2(1-36 F_\pi^2 \overline{C}_T)+4 g_A^4}{96 F_\pi^4\pi^2}\;.
\eeqa
Performing the same shifts in the scalar current and
using $L(0)=1$ and Eq.~(\ref{three_point_f_at_0}) we indeed verify:
\beqa
S_{\rm 2N}^{(Q^{-2})}(k=0)+S_{\rm 2N}^{(Q^{0})}(k=0)&=&\frac{\partial V^{\rm OPE}}{\partial
  m_q}\;,\nn
S_{\rm 2N:2\pi}^{(Q^{0})}(k=0)+S_{\rm 2N: cont}^{(Q^{0})}(k=0)&=&\frac{\partial}{\partial
  m_q}\Big(V^{\rm TPE}+V^{\rm cont}+\delta V^{\rm
  cont}\Big)\;.
\eeqa

\end{document}